\documentclass{elsart}
\usepackage{graphicx}
\newcommand{\<}{\langle}
\renewcommand{\>}{\rangle}
\begin{document}
\begin{frontmatter}
\title{Local adiabatic quantum search with different paths}
\author[ankara]{Recep Eryi\u{g}it},
\author[ankara]{Yi\u{g}it G\"{u}nd\"{u}\c{c}} and
\author[bolu]{Resul Eryi\u{g}it\thanksref{email}}
\address[ankara]{Department of Engineering Physics, Hacettepe University,
Ankara, 06120-Turkey.}
\address[bolu]{Department of
Physics, Abant Izzet Baysal University, Bolu, 14280-Turkey.}
\thanks[email]{email:resul@ibu.edu.tr}

\begin{abstract}
We report on a detailed analysis of generalization of the local
adiabatic search algorithm. Instead of evolving directly from an
initial ground state Hamiltonian to a solution Hamiltonian a
different evolution path via a 'sure success' Hamiltonian is
introduced and is shown that the time required to find an item in
a database of size $N$ can be made to be almost independent of the
size of the database, asymptotically by modifying the Hamiltonian
used to evolve the system. The complexity of the search algorithm
in this and other similar approaches is shifted to construction of
the Hamiltonian.
\end{abstract}
\begin{keyword}
Quantum search algorithms; Local adiabatic search algorithm
\end{keyword}
 \end{frontmatter}
\section{Introduction}
One of the promising approaches to quantum computing is provided
by the adiabatic evolution method introduced by Farhi \textit{et.
al.}\cite{farhi98} and elaborated and developed by several
groups~\cite{farhi98,das03,roland02}. While the standard paradigm
of quantum computation uses quantum gates (i.e., unitary
operators) applied to quantum registers, the main idea of the
adiabatic computing algorithms is to encode the initial state and
solution state of the problem at hand in properly chosen
Hamiltonians. Generally, the meaning of the "proper" is the
requirement that the initial and the solution states are the
ground states of an initial and final Hamiltonians. Then these two
Hamiltonians are combined in a time-dependent way and system is
let to evolve under the influence of this combined Hamiltonian.
The speed of mixing is determined by the adiabatic theorem of
quantum mechanics~\cite{kato50,messiah61}. According to this
theorem, if the evolution of a quantum system is governed by a
Hamiltonian that varies slowly enough, the system will stay near
its instantaneous ground state. The general time-complexity of
adiabatic quantum computation is still an open
problem~\cite{kaminsky02}. The complexity of the algorithm is
determined by the time needed for switching from initial to
solution state "slowly enough". The switching speed depends
inversely on the size of the spectral gap between the
instantaneous ground and first excited states of the Hamiltonian.

 One of the successful quantum algorithms is the
database search algorithm of Grover~\cite{grover97}, which has
been examined in great details in many
studies~\cite{zalka97,boyer98,biham99}. This algorithm shows a
quadratic speedup in searching a database over classical
algorithms. Adiabatic search algorithm in its local adiabatic
version was shown to provide similar speedup in searching, namely
$O(\sqrt{N})$ number of queries~\cite{roland02}, where $N$ is the
number of items in the database. Also, some numerical evidence
suggests that adiabatic algorithm might efficiently solve some
instances of hard combinatorial problems, outperforming classical
methods~\cite{farhi00,childs00,farhi01,childs01}.

Farhi \textit{et. al.}~\cite{farhi02} showed that one does not
need to take a straight path from initial to final Hamiltonian.
They showed that by applying a control Hamiltonain at the
evolution stage an algorithmic failure can be turned to a success
for 3-SAT problem. Recently, Boulatov and Smelyanskiy
reported~\cite{boulatov03} a detailed theoretical study of using
different paths for a version of 3-SAT problem.

In the present article, we show that one can get a better than
square-root speedup by using different path approach of
Ref.~\cite{farhi02}. We use part of 'sure success' Hamiltonian of
Bae and Kwon~\cite{bae02} as the control Hamiltonian. Added
control Hamiltonian keeps the minimum spectral gap a constant plus
a system size dependent term which goes to zero as the system size
tends to get close to infinity. It is shown that complexity of the
proposed algorithm approaches a constant almost independent of the
size of the system.

\section{Quantum Adiabatic Theorem}

Consider a quantum system in state $\left|\Psi(t)\right>$ which
evolves according to the  Schrodinger equation

\begin{eqnarray*}
i\frac{d}{dt}\left|\Psi(t)\right>=H(t)\left|\Psi(t)\right>
\end{eqnarray*}

\noindent where $H(t)$ is the Hamiltonian governing the dynamics
of the system (we take $\hbar=1$). If the Hamiltonian is
independent of time and the system is at its ground state, it will
remain there forever. According to adiabatic
theorem~\cite{kato50,messiah61} if the Hamiltonian varies slowly
enough, the state of the system will remain at the instantaneous
ground state of the Hamiltonian for all $t$.

Let $\left|\Psi_{k}(t)\right>$ be the eigenstate of $H(t)$, which
satisfy
\begin{eqnarray*}
H(t)\left|\Psi_{k}(t)\right>=E_{k}(t)\left|\Psi_{k}(t)\right>
\end{eqnarray*}
\noindent where $E_{k}(t)$ is the corresponding eigenvalue and k
labels the eigenvalues ($k=0$ is the ground state).

 The adiabacity conditions are related to the minimum energy gap
of the spectrum of $H(t)$ and the transition to excited states of
the system. These conditions can be made as follows: The minimum
energy gap is defined as
\begin{eqnarray*}
g_{\rm{min}}=\min_{0\le t\le T}{\left[E_{1}(t)-E_{0}(t)\right]}
\end{eqnarray*}
\noindent and the maximum of the matrix element of $dH/dt$ between
the ground and the first excited state;
\begin{eqnarray}
D_{\rm{max}}=\max_{0\le t\le
T}{\left|\left<\Psi_{1}(t)\right|\frac{dH}{dt}\left|\Psi_{0}(t)\right>\right|}
\label{dmax}
\end{eqnarray}

Adiabatic theorem states that, if we prepare the system at time
$t=0$ in its ground state $\left|\Psi_{0}(0)\right>$ and let it
evolve under the Hamiltonian $H(t)$ for a time $T$, then
\begin{eqnarray*}
\left|\left< \Psi_{0}(T)\right. \left|\Psi(T)\right>\right|^2 \ge
1-\epsilon^2
\end{eqnarray*}
\noindent provided that
\begin{eqnarray}
\frac{D_{max}}{g_{min}^{2}}\le \epsilon \label{adcon}
\end{eqnarray}
\noindent where $\epsilon \ll 1$.

\section{Local Adiabatic Evolution Algorithm}
The problem we consider here is to find a marked item in a set of
$N$ items. Items in the system are labeled by $n$ qubits, so the
dimension of the Hilbert space of the system is $N=2^n$. The
initial state of the system is the uniform superposition of all
basis states, which also includes the unknown marked state $|m\>$

\begin{eqnarray}
|\psi_{0}\>=\frac{1}{\sqrt{N}}\sum_{i=0}^{N-1}|i\> \label{initial}
\end{eqnarray}
\noindent where $|i\>$ are the basis states with $i=0,\ldots,N-1$.
$|\psi_{0}\>$ is the zero energy ground state of
\begin{eqnarray*}
H_{0}=I-\left|\psi_{0}\>\<\psi_{0}\right|
\end{eqnarray*}
whose ground state is $|\psi_{0}\>$ with energy $0$. Let the
marked state Hamiltonian be
\begin{eqnarray*}
H_{m}=I-\left|m\>\<m\right|
\end{eqnarray*}
\noindent whose ground state, the marked state $|m\>$, is unknown.
The time-dependent Hamiltonian that interpolates between these two
Hamiltonians is
\begin{eqnarray}
H(s)=(1-s)H_{0}+sH_{m}
\label{tdh1}
\end{eqnarray}
\noindent where the mixing parameter $s=s(t)$ is a monotonic function with
$s(0)=0$ and $s(T)=1$, $T$ being the required running time of the algorithm. The algorithm
is to prepare the system in the state $|\psi(0)\>=|\psi_{0}\>$ and
then apply $H(s)$ for a time interval $T$ to end up with the
state $|m\>$. Roland and Cerf~\cite{roland02} shown that
complexity of this algorithm when applied in local adiabatic
scheme scales with $\sqrt{N}$, similar to Grover's algorithm and
this result is optimal for any chose of the time evolution of
$s(t)$.

\begin{figure}[htb]
\centering
\includegraphics[width=8cm]{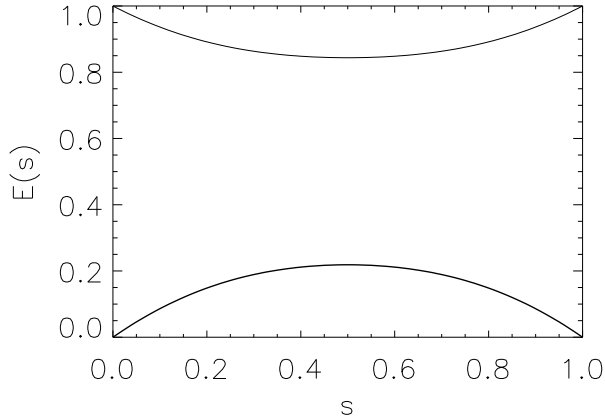}
\caption{Eigenvalues of the two lowest states of the
time-dependent Hamiltonian $\tilde{H}(s)$ as a function of reduced
time $s$ for $N=64$.} \label{figeigen}
\end{figure}

One can generalize the time-dependent Hamiltonian of
Eq.~\ref{tdh1} by adding to it a Hamiltonian of the
form~\cite{fenner00,bae02}
\begin{eqnarray*}
H_{D}=a(s)|\psi_{0}\>\<m|+b(s)|m\>\<\psi_{0}|
\end{eqnarray*}
\noindent where we have chosen $a(s)=b(s)=s(1-s)$ so that the
Hamiltonian at $s=0$ is $H_{0}$ and at $s=1$ it is $H_{m}$.
$H_{D}$ acts only during the adiabatic evolution of the system. So
its effect is to change the path taken by the system for going
from the initial state $|\psi_{0}\>$ to final state $|m\>$. A
similar extra Hamiltonian was shown to convert an algorithmic
failure to a success by Farhi \textit{et.al.}~\cite{farhi02}. Now the
time-dependent Hamiltonian of the system is
\begin{eqnarray}
H(s)=(1-s)H_{0}+sH_{m}-H_{D} \label{tdh2}
\end{eqnarray}
\noindent which in matrix representation can be written as
\begin{eqnarray*}
H=E\left(\begin{array}{cc}

(s-1)(x(s+x)-1) & (s-1)(s+x)\sqrt{1-x^2}\\

(s-1)(s+x)\sqrt{1-x^2}&s(1-s)x^2\end{array}\right)
\end{eqnarray*}
\noindent where $x=\<m|\psi_0\>=1/\sqrt{N}$. Two lowest eigenvalues
of this Hamiltonian are plotted as function of $s$ in
Fig.~\ref{figeigen}. These eigenvalues are separated by a gap
\begin{eqnarray}
g(s)=\sqrt{1-4s(1-s)\left(1-x(x+1)-s(1-s)\right)} \label{gap}
\end{eqnarray}
\noindent note that the minimum of $g(s)$ is at $s=1/2$ with value
$g_{\rm{min}}=1/2+1/\sqrt{N}$.

The eigenvectors corresponding to two lowest eigenvalues are
\begin{eqnarray*}
|E_{0};s\>&=&\frac{(1 -g(s) - 2\,s)|m>+\left( s-1 \right) \,\left(
s + x \right) \,\left( 1 + {\sqrt{1 - x^2}}
\right)|\psi_0>}{\sqrt{4\,{\left( s-1 \right) }^2\,{\left( s + x
\right) }^2\,\left( 1 - x^2 \right)  +
       {\left( 1 -g(s) - 2\,s + 2\left(s-1\right)x\left(s+x\right)
           \right) }^2}}\\
|E_{1};s\>&=&\frac{(1 + g(s) - 2\,s)|m>+\left( s-1 \right)
\,\left( s + x \right) \,\left( 1 + {\sqrt{1 - x^2}}
\right)|\psi_0>}{\sqrt{4\,{\left( s-1 \right) }^2\,{\left( s + x
\right) }^2\,\left( 1 - x^2 \right)  +
       {\left( 1 +g(s) - 2\,s + 2\left(s-1\right)x\left(s+x\right)
           \right) }^2}}
\end{eqnarray*}

The matrix element in Eq.~\ref{dmax} can be written as
\begin{eqnarray}
\left\<\frac{dH}{dt}\right\>_{1,0}&=&\frac{ds}{dt}\left\<\frac{d\tilde{H}}{ds}\right\>_{1,0} \nonumber\\
&=&\frac{ds}{dt}(\left\<
H_{m}-H_{0}+(1-2s)H_{D}\right\>_{1,0})\label{dht}
\end{eqnarray}
\noindent $\left\<
H_{m}-H_{0}+(1-2s)H_{D}\right\>_{1,0}\le 1$~\cite{roland02}.

\begin{figure}[htb]
\centering
\includegraphics[width=8cm]{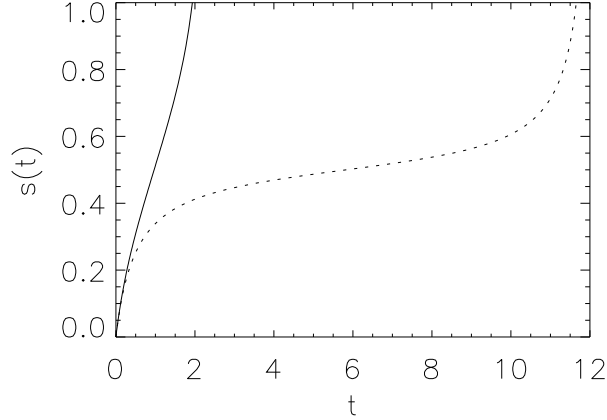}
\caption{Dynamic evolution of the Hamiltonian that drives the
initial ground state to the solution state: plot of the evolution
function $s(t)$ for $N=64$. The dotted line is for the same
quantity calculated by using the Hamiltonian of
Ref.~\cite{roland02}.} \label{figsvt}
\end{figure}

In the the local adiabatic approximation one tries to adjust the
evolution rate of the $s$ for infinitesimal time intervals $dt$
and apply the adiabacity condition locally to each of these
intervals. The adiabacity condition for the local adiabatic
evolution would be

\begin{eqnarray*}
\left|\frac{ds}{dt}\right| \le \epsilon
\frac{g^{2}(s)}{\left|\left\< \frac{d\tilde{H}}{ds} \right\>
_{1,0}\right|}
\end{eqnarray*}
\noindent for all times $t$. Using Eq.~\ref{gap}, the evolution of
the Hamiltonian is chosen to evolve at a rate that is solution of
\begin{eqnarray}
\frac{ds}{dt}=\epsilon g^{2}(s)=\epsilon\left(
1-4s(1-s)\left(1-\frac{1}{N}-\frac{1}{\sqrt{N}}+s(s-1)\right)\right)
\label{srate}
\end{eqnarray}
\noindent Integral of this equation involves the logs of the roots
of $g^2(s)$ and is displayed in Eq.~\ref{integral}. $s(t)$
obtained from this integral is shown in Fig.~\ref{figsvt} along
with the $s(t)$ of Roland and Cerf's algorithm.
\begin{eqnarray}
t=\frac{1}{\sqrt{2}h(N)}\left(\frac{1}{k_2(N)}\tanh^{-1}{\left[\frac{2\sqrt{2}k_2(N)s}{2-k_2(N)^2-4s}\right]}-
\frac{1}{k_1(N)}\tanh^{-1}{\left[\frac{2\sqrt{2}k_1(N)s}{2-k_1(N)^2-4s}\right]}\right)
\label{integral}
\end{eqnarray}
\noindent where
\begin{eqnarray*}
h(N)&=&\frac{1}{N}\sqrt{(1-N)(2\sqrt{N}+1)} \\
k_{1}(N)&=&\sqrt{-2-4 h(N)+\frac{4}{N}+\frac{4}{\sqrt{N}}} \\
k_{2}(N)&=&\sqrt{-2+4 h(N)+\frac{4}{N}+\frac{4}{\sqrt{N}}}
\end{eqnarray*}

\begin{figure}[htb]
\centering
\includegraphics[width=8cm]{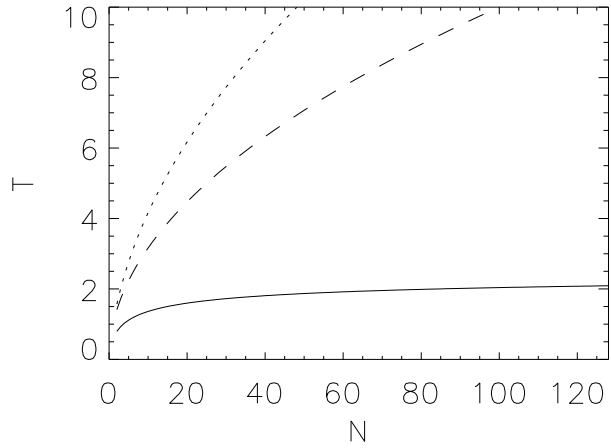}
\caption{Computation time vs the size of the problem. The dashed
and dotted line is for the $\sqrt{N}$ and Hamiltonian of
Ref.~\cite{roland02}, respectively. } \label{figtime}
\end{figure}

As can be seen from the figure, added term $H_{D}$  changes $H(s)$
much faster compared to the original Hamiltonian of
Ref.~\cite{roland02}. The computation time of the algorithm is
evaluated by taking $s=1$ in the solution of Eq.~\ref{srate},
 and is plotted against $N$ in Fig.~\ref{figtime} along
with the same quantity from Ref.~\cite{roland02}. The asymptotic
form of Eq.~\ref{srate} for $s=1$ is given by
\begin{eqnarray}
T=
\left(1+\frac{\pi}{4}\right)+O\left(1/\sqrt{N}\right)\label{compT}
\end{eqnarray}
\noindent which show that as the system size $N$ approaches infinity, the required
running time approaches the constant $\left(1+\frac{\pi}{4}\right)$.

The main factor that determines the required evolution time $T$ is
the behavior of minimum gap $g_{\rm{min}}$. This energy difference
usually reaches its minimum at an avoided level-crossing. If this
approach is exponentially fast with a growing particle number then
the algorithm is not efficient. The effect of extra term $H_{D}$
is to keep these two levels far apart. Our result seem to be
similar to one obtained by Das \textit{et. al.}~\cite{das03} who
use Eq.~\ref{tdh1} with different coefficients for $H_{0}$ and
$H_{m}$ terms. One can, actually have a minimum gap that is larger than Eq.~\ref{gap}
 by choosing $a(s)=b(s)=\sqrt{s(s+1)}$ in the definition of
 the driving Hamiltonian $H_{D}$. In that case, the running time approaches $1$ for
 the large $N$ limit.

\section{Conclusion}
In summary, we have presented
the results of an analysis of a generalized adiabatic quantum
search algorithm. Our analysis shows that the optimal speed
associated with Grover's search algorithm can be improved by a
careful choice of the time-dependant Hamiltonian. This analysis does not
deal with the complexity of implementing the driving Hamiltonian. Hamiltonian of
adiabatic algorithms  includes searched state $|m\>$ terms. The construction of
these terms are thought to be analogous to having access to an oracle in original
Grover algorithm. It seems that the time-complexity of searching in present approach is
shifted from searching process to construction of the Hamiltonian.

\end{document}